# Gas flow and accretion via spiral streamers and circumstellar disks in a young binary protostar

**Authors:** F. O. Alves[1,*], P. Caselli[1], J. M. Girart[2,3], D. Segura-Cox[1], G. A. P. Franco[4], A. Schmiedeke[1], B. Zhao[1]

**Affiliations:**

[1] Center for Astrochemical Studies, Max Planck Institute for Extraterrestrial Physics, Garching, 85748, Germany

[2] Institut de Ciències de l'Espai, Consejo Superior de Investigaciones Científicas, Cerdanyola del Vallès, E-08193, Catalunya, Spain

[3] Institut d'Estudis Espacials de Catalunya. Barcelona, E-08034, Spain

[4] Departamento de Física, Instituto de Ciências Exatas, Universidade Federal de Minas Gerais, Belo Horizonte, 30.123-970, Brazil

*Correspondence to: falves@mpe.mpg.de

The majority of stars are part of gravitationally bound stellar systems, such as binaries. Observations of proto-binary systems constrain the conditions that lead to stellar multiplicity and subsequent orbital evolution. We report high angular resolution observations of the circumbinary disk around [BHB2007] 11, a young binary protostar system. The two protostars are embedded in circumstellar disks with radii of 2-3 astronomical units, probably containing a few Jupiter masses. These systems are surrounded by a complex structure of filaments connecting to the larger circumbinary disk. We also observe accretion and radio jets associated with the protobinary system. The accretion is preferentially onto the lower-mass protostar, consistent with theoretical predictions.

**One sentence summary**: High-resolution images show a young stellar binary system nurtured by a circumbinary disk through a complex network of accretion filaments

About half the stars in the solar neighborhood are in gravitationally bound stellar systems such as binaries (*1*). The mean separation between components, a few tens of astronomical units (au, 1 au = 149,597,870.7 km) is a consequence of their formation process, thought to be fragmentation of the protostellar disk due to gravitational instabilities (*2, 3*). Observations of young binary systems rarely probe such small scales. We seek to investigate the dynamical evolution of a binary system still embedded in its natal cloud, where gas from a circumbinary disk is expected to accrete onto each binary component. In the protobinary accretion phase, theoretical models indicate that high angular momentum material accretes preferentially onto the less-massive companion, which has a higher orbital angular momentum than the primary, causing ultimately an equal share of mass among the individual components (*4-6*).

[BHB2007] 11 ($\alpha$ = $17^h11^m23.18^s$, $\delta$ = -27°24′31.5″)[1] is the youngest member (age 0.1−0.2 Myr, (*7*)) of the small cluster of young stellar objects in the Barnard 59 core (B59, part of the Pipe Nebula molecular cloud), which is still growing mass through dust and gas accretion (a Class 0/I young stellar object). Previous observations of this object show an envelope surrounding a circumbinary disk of radius 90 au, with prominent bipolar outflows launched near the disk edge (*8*).

We observed [BHB2007] 11 at 225.3 GHz (wavelength $\lambda_0$ ~ 1.3 mm) with the Atacama Large Millimeter/submillimeter Array (ALMA) (*9*). The observations were centered on the circumbinary disk of [BHB2007] 11 and reveal its internal structure. Figure 1 shows a multi-scale view of the dust distribution from the B59 core scale (~20,600 au) to sub-disk scale (~6 au), the latter is the maximum spatial resolution of our map given the B59 distance of 163 pc (*10*). The dust map reveals two compact sources that we interpret as circumstellar disks around

---
[1] Coordinates in Julian Epoch 2000 as found in the SIMBAD Astronomical Database

both components of a protobinary system, which has a projected separation of ~28 au. The components were also seen in previous observations (*11, 12*) using the Karl G. Jansky Very Large Array (VLA). We design the individual components as [BHB2007] 11A (northern source) and [BHB2007] 11B (southern source). The dust emission seen in the ALMA data arises from circumstellar disks with radii 3.1 ± 0.6 au ([BHB2007] 11A) and 2.1 ± 0.5 au ([BHB2007] 11B), similar to the radius of the asteroid belt in the Solar System. Their inclinations are ~ 40° between the disk normals and the line-of-sight. From the continuum emission, the masses of the circumstellar disks are estimated to be of the order of a Jupiter mass ($M_{Jup}$), with the [BHB2007] 11A disk slightly more massive than the [BHB2007] 11B disk (*9*). The dust content in the disks is of a few Earth Masses ($M_{Earth}$). The larger circumbinary disk has a total mass of 0.08 ± 0.03 solar masses ($M_\odot$, where 1 $M_\odot$ ~ 1000 $M_{Jup}$ ~ 330000 $M_{Earth}$). Because this represents ~260 Earth masses in dust, we speculate that the circumstellar disks may form rocky terrestrial planets (*13*).

The protobinary is in the center of a complex network of dust structures distributed in spiral shapes. Spirals in protostellar systems such as LDN 1448NB have been interpreted as the outcome of disk fragmentation (*14*). The Toomre parameter, which refers to the dynamical state of the disk (*15*), shows that the circumbinary disk in [BHB2007] 11 is stable against gravity (*9*). Therefore the observed spirals unlikely result from disk fragmentation. The total length of the filaments is ~ 392 au, their mean width is ~ 12 au and the mean brightness temperature $T_b$ is 14 K, from which we estimate a mean total mass of ~19 $M_{Jup}$. (*9*). This corresponds to a mean mass per unit length of ~10 $M_\odot$ pc$^{-1}$ (1 pc = 206,264.8 au), which is below the critical value required to produce fragmentation in the filaments. The spiral shape centered on the protobinary system indicates dynamical interaction with the protostars. Figure 1B shows that the geometrical center of the circumbinary disk is located near [BHB2007] 11A, within the angular resolution, suggesting that this source is the primary, higher mass object of the system. We suggest that the filaments are inflow streamers from the extended circumbinary disk onto the circumstellar disks of the protobinary system. We estimate a mass accretion rate from the circumbinary disk into the circumstellar disk of ~ $1.1 \times 10^{-5}$ $M_\odot$ year$^{-1}$ (*9*), which is consistent with other protostellar sources (*16*). The rotationally supported disk may be redistributing its angular momentum (through processes such as viscosity, magnetic braking or gravitational torques), causing material to fall into the gravitational potential wells of the individual protostars. While other factors such as envelope accretion into the circumbinary disk, misaligned disks or eccentric binary orbit could also produce substructure such as filaments in the circumbinary disk (*6, 17, 18*), accretion streamers feeding the circumstellar disks can still happen in those scenarios (*6, 17*).

We measured kinematics by re-analyzing (*9*) previous observations (*8*) of the CO (*J*=2→1) molecular emission line in [BHB2007] 11. The relative positions of the molecular gas in each velocity channel of the emission spectra are determined with positional accuracies better than 0.05″ (*9*). The systemic velocity of [BHB2007] 11 with respect to the local standard of rest is $v_{lsr}$ ~ 3.6 km s$^{-1}$. The CO line traces primarily outflow emission but velocity components faster than the outflows are detected within the circumbinary disk (*8*). These high-velocity components reach up to 15 km s$^{-1}$. The blue and redshifted velocities ($v_{lsr}$ < -1.5 km s$^{-1}$ and $v_{lsr}$ > 9 km s$^{-1}$, respectively) are centered on [BHB2007] 11B, with the higher velocities closest to the position of the protostar (Figure 2A). No high-velocity components are detected near [BHB2007] 11A. The increasing velocity toward [BHB2007] 11B indicates acceleration of infalling gas from the circumbinary disk onto [BHB2007] 11B. The observed velocities are consistent with the filament accretion rate estimated above. Figure 2B shows the CO spectrum with the high-velocity channels used to compute the CO emission peaks.

We used the VLA to observe the protobinary in continuum emission at 34.5 GHz (*11*) and at lower frequencies (10, 15 and 22 GHz) (*9*). The centimeter-wavelength data are consistent with thermal free-free emission from partially ionized collimated jets produced as an outcome of the angular momentum redistribution in the disk-star system. This type of emission correlates with the momentum rate, so the radio flux is also correlated with the disk-star accretion rate (*19*). The radio jet-like emission along the East-West direction associated with [BHB2007] 11A (Figure 3) suggests that this source has higher (circumstellar) disk-star accretion rate than [BHB2007] 11B. In addition, [BHB2007] 11A has an ionized mass loss rate a factor of ~2 higher than [BHB2007] 11B (~$6.4 \times 10^{-10}$ $M_\odot$ year$^{-1}$ and ~$3.7 \times 10^{-10}$ $M_\odot$ year$^{-1}$, respectively, (*9*)). The stronger radio jet and higher ionized mass loss rate support the proposition that [BHB2007] 11A is the primary source of the system. The infalling gas onto [BHB2007] 11B suggests a higher accretion from the circumbinary disk onto this object than [BHB2007] 11A. This preferential accretion onto a companion stellar object in a binary system matches theoretical predictions (*4-6*).

In summary, the small-scale structure seen in our observations includes streamers of gas and dust accreting from a circumbinary disk onto small circumstellar disks around the individual components of a protobinary system. Their disks have a dust mass equivalent to a few Earth masses. The gas accretion is more prominent in the secondary, less massive object of the system, while the disk-star accretion inferred by the ionized free-free emission is higher in the primary object instead. We expect this two-level accretion process, circumbinary disk to circumstellar disks then circumstellar disks to stars, drives the dynamics of the binary system during its mass accretion phase.

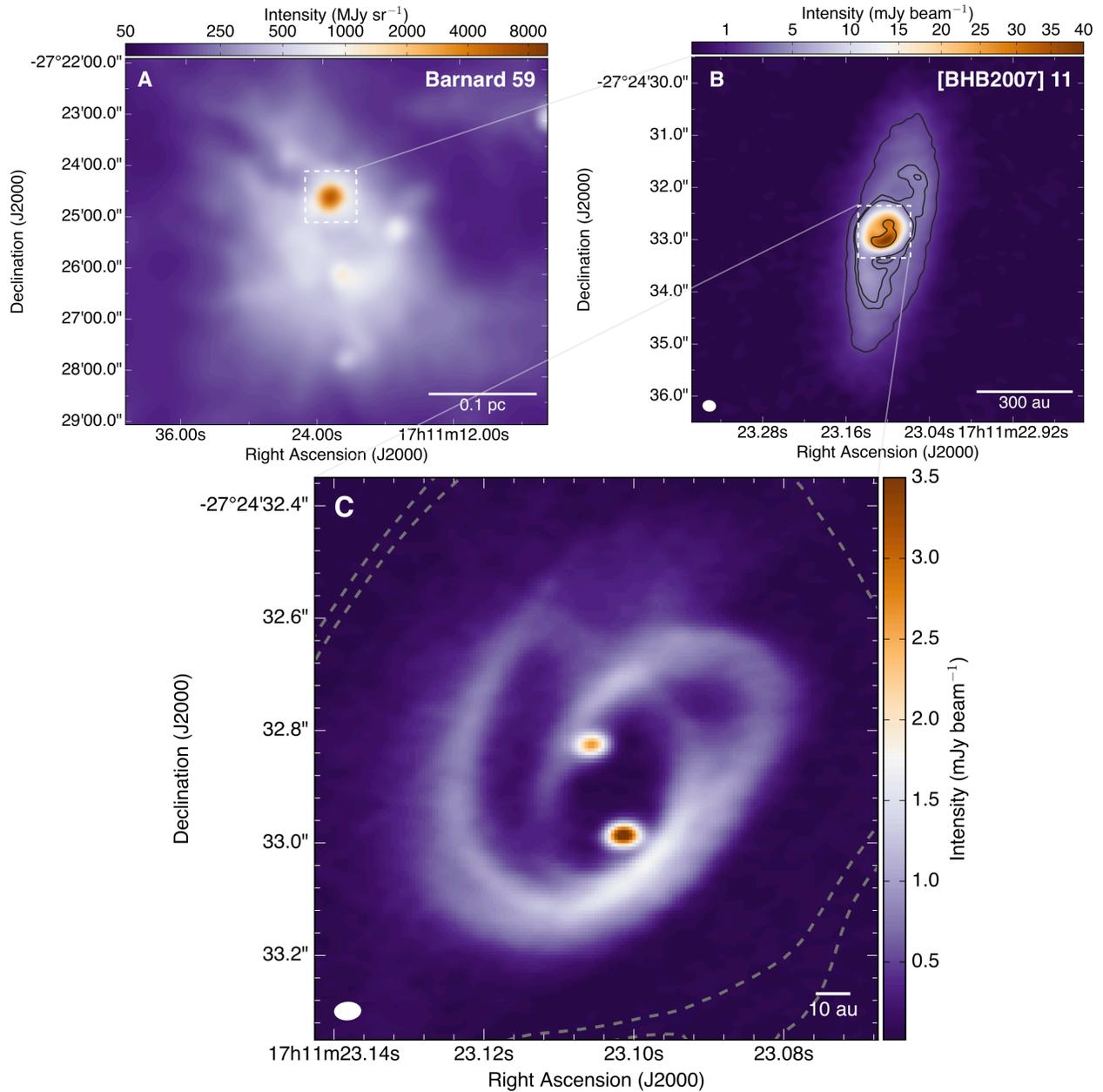

**Fig. 1. The dust distribution in [BHB2007] 11 on scales from the core to the circumstellar disks.** (**A**) The core of Barnard 59 observed at 250 μm with the *Herschel* space telescope (*20*). (**B**) The [BHB2007] 11 disk and envelope observed at 1.3 mm with ALMA (*8*). The contours indicate the 30, 60, 70, 450 intensity levels times the map noise of 62 μJy beam$^{-1}$. Intensity levels are scaled logarithmically in panels A and B. (**C**) A zoom into the circumbinary disk, revealing the dust filaments surrounding the protobinary system, where [BHB2007] 11A is the northern

component and [BHB2007] 11B is the southern and brighter component. The dashed lines are the intensity contours from panel (B). Panels B and C show the ALMA synthesized in the bottom left corner.

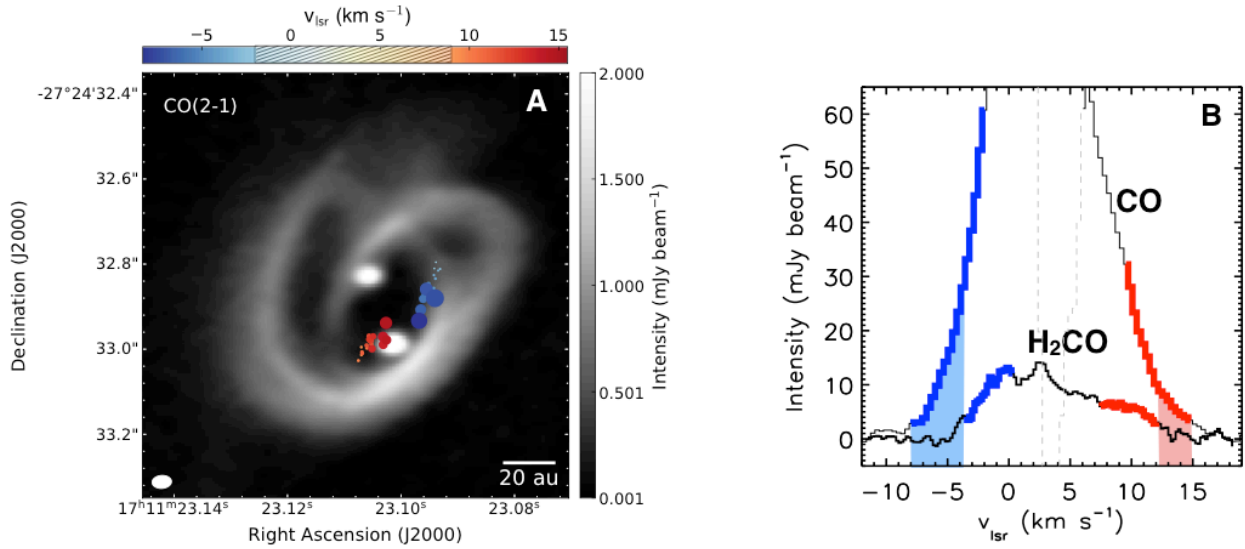

**Fig. 2. Small-scale kinematics in [BHB2007] 11 measured from CO molecular line emission.** **(A)** High velocity components of the CO emission (with respect to the ambient local standard of rest velocity of [BHB2007] 11, $v_{lsr} \sim 3.6$ km s$^{-1}$) overlaid on the continuum map (grey scale). Circles indicate the position of CO emission peaks in each velocity channel (color bar). CO data between -2 km s$^{-1}$ and 9 km s$^{-1}$ (indicated as a shaded area in the color bar) are not shown because most gas close to the ambient velocity are filtered out by the interferometer and/or trace outflow/envelope extended emission. Circle sizes equal the uncertainty on the absolute position (*9*). **(B)** CO mean spectrum extracted from the disk area showing the CO velocity channels used in (A). The shaded areas indicate the CO velocity components that exceed the reference H$_2$CO spectrum associated with the rotation of the circumbinary disk (*9*). The faint dashed line shows the strong filtering of the CO line near the systemic velocity of the source.

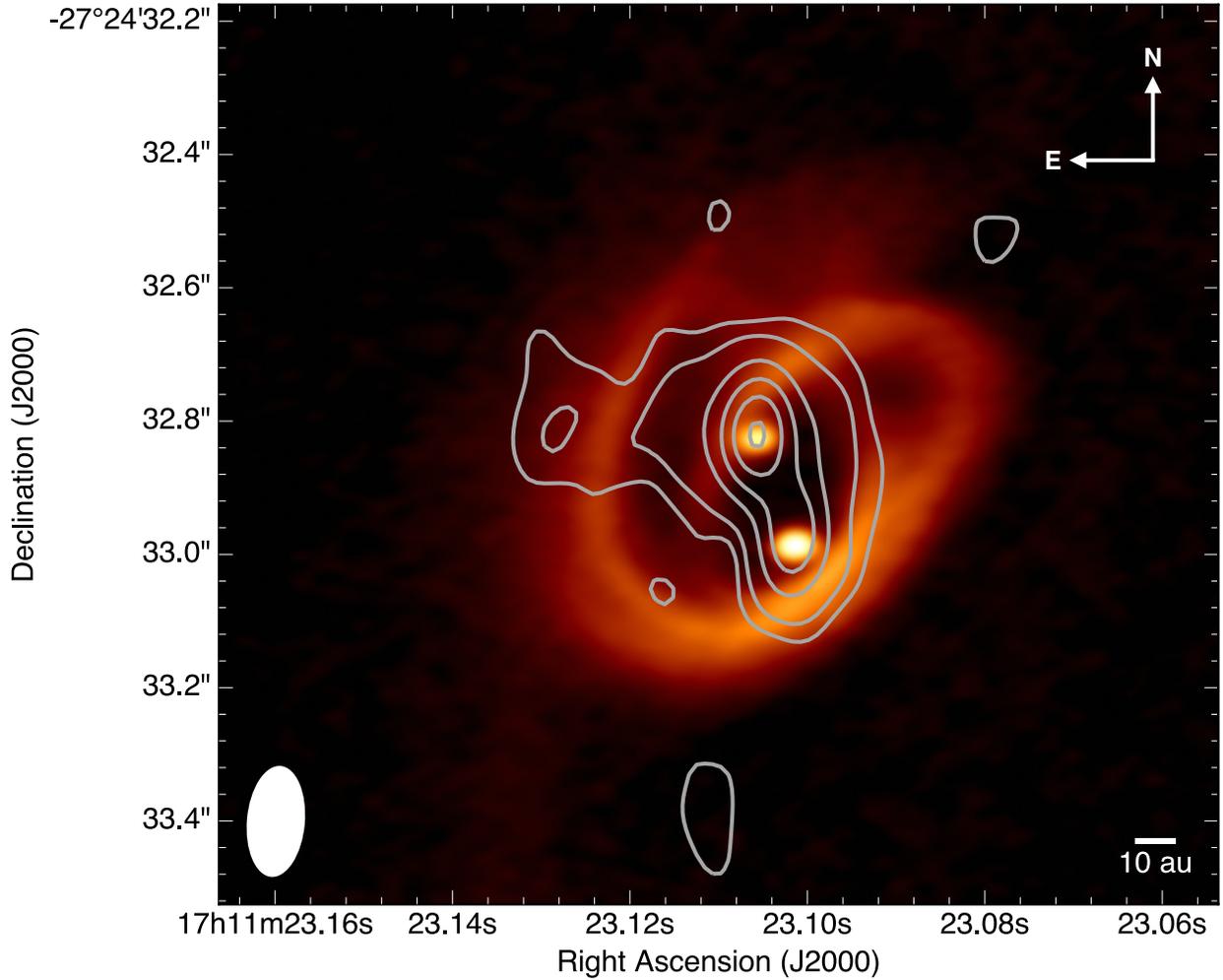

**Fig. 3. Centimetric emission associated with the binary system.** Continuum emission at 0.94 cm (32.5 GHz) associated with the [BHB2007] 11 binary system, observed with the VLA. Contours represent the 3, 5, 10, 15, 20, 25 times the noise of the map, 12 μJy beam$^{-1}$; the background image is the ALMA map from Figure 1A. The extended emission to the east of [BHB2007] 11A suggests a radio jet associated with this source. The VLA synthesized beam is shown in the lower left corner.

**Acknowledgements**: We thank the reviewers for helpful and detailed comments that improved the manuscript and the ALMA and VLA staff for performing the observations and quality assessment of the data. We also thank Manual Fernández López, Wladmir Lyra and Paulo Cortes for useful discussions on the data analysis and interpretation. ALMA is a partnership of ESO (representing its member states), NSF (USA) and NINS (Japan), together with NRC (Canada), MOST and ASIAA (Taiwan), and KASI (Republic of Korea), in cooperation with the Republic of Chile. The Joint ALMA Observatory is operated by ESO, AUI/NRAO and NAOJ. The VLA is an instrument of the National Radio Astronomy Observatory, a facility of the National Science Foundation operated under cooperative agreement by Associated Universities, Inc. This research has also made use of data from the Herschel Gould Belt survey (HGBS) project (http://gouldbelt-herschel.cea.fr). The HGBS is a Herschel Key Programme jointly carried out by SPIRE Specialist Astronomy Group 3 (SAG 3), scientists of several institutes in the PACS Consortium (CEA Saclay, INAF-IFSI Rome and INAF-Arcetri, KU Leuven, MPIA Heidelberg), and scientists of the Herschel Science Center (HSC).

**Funding**: F.O.A., P.C., D.S.-C, A.S. and B.Z. acknowledge financial support from the Max Planck Society. P.C. and B.Z. acknowledge support of the European Research Council (ERC, project PALs 320620). J.M.G. is supported by the MINECO (Spain) grant AYA2017-84390-C2. G.A.P.F. acknowledges support from CNPq and FAPEMIG (Brazil).

**Author contributions**: F.O.A. led the project and ALMA observing proposal, analyzed the ALMA data (calibration and imaging), conducted the molecular line data analysis and led the preparation of the manuscript; P.C. contributed to paper preparation, molecular line interpretation and derivation of dust properties; J.M.G. led the VLA observing proposal and the VLA data analysis, including calibration, imaging and spectral index derivation; D.S.C. performed the self-calibration of the continuum image and contributed to the continuum data analysis; G.A.P.F. developed the analytical approach for the mass accretion rate and contributed to the manuscript preparation; A.S. and B.Z. contributed to the theoretical interpretation of the results. All authors contributed to the discussion and interpretation of the data.

**Competing interests**: There are no conflicts of interests.

**Data availability**: The data reported in this paper are archived in the ALMA Science Archive http://almascience.nrao.edu/asax/ under project codes ADS/JAO.ALMA#2016.1.01186.S and ADS/JAO.ALMA#2013.1.00291.S. The VLA data are available from https://science.nrao.edu/facilities/vla/archive/index under project VLA/16B-290.

**Supplementary Materials**

Materials and Methods

Table S1

Fig S1, S2, S3

References (21 – 29)

# Supplementary Materials for

## Gas flow and accretion via spiral streamers and circumstellar disks in a young binary protostar

F. O. Alves, P. Caselli, J. M. Girart, D. Segura-Cox, G. A. P. Franco, A. Schmiedeke, B. Zhao

Correspondence to: falves@mpe.mpg.de

**This PDF file includes:**

    Materials and Methods
    Figs. S1 to S3
    Table S1

**Materials and Methods**

Observations
 The Band 6 (225.3 GHz) ALMA data were observed in two Execution Blocks (EBs) on the 05$^{th}$ and 23$^{rd}$ of November 2017, with 43 and 48 antennas of the main array, respectively. The total on-source observing time was 2.4 hours and the mean precipitable water vapor was 0.5 and 1.4 mm $H_2O$ for the first and second EB, respectively. The antenna configurations had baselines between 92.1 m and 13.9 km. This implies that the observations are sensitive to an extended emission of ~ 0.8″ (~ 130 au, at the B59 distance of 163 pc (*10*)), with the larger scales being spatially filtered out by the array.
 Bandpass calibration of the receiver response was performed with observations of QSO B1921-293. Bandpass shapes are nearly flat with variations of less than 5% in amplitude gains and less than 10° in phase gains across the bandwidth for most antennas. Absolute flux calibration was obtained by bootstrapping of the flux density of QSO B1730-130 to the bandpass calibrator, QSO B1921-293, gain calibrator, QSO J1700-2610, and the target [BHB2007] 11. We expect an uncertainty of about 15% in the absolute flux density scale to account for flux variability of the calibrator, spectral index uncertainty from ALMA monitoring observations in other frequency bands, and inaccuracies in the models of solar system objects used as primary amplitude calibrators by ALMA.
 The standard calibration of the visibilities was performed by the ALMA staff as part of the Quality Assessment (QA2) process. The full continuum bandwidth is 2.4 GHz. We have self-calibrated the continuum using CASA 5.4.0 (*21*) with decreasing solution intervals with a shortest phase cycle equal to the integration length. We then performed one round of self-calibration on the amplitude gains with one solution per observation execution. We produced maps with the CASA tclean task using Briggs weighting (robust factor 0.5), and multiscale cleaning on scales of 0, 1, 5 and 15 times the size of the synthesized beam to recover extended flux. The final map after self-calibration shows an increase in signal-to-noise ratio (SNR) of about 13% and improvement in image fidelity, resulting in enhanced sensitivity to low-amplitude extended structures with respect to the QA2 products. The final continuum image has an angular resolution (synthesized beam) of 45 × 30 milliarcseconds (mas) (~ 6 au) and a position angle of −86°. The noise level of the image is 19 µJy beam$^{-1}$.
 The observing setup of the ALMA Band 3 (97.5 GHz) continuum data discussed only in this supplementary section was reported in (*11*), which presented images obtained using part of the observed visibilities (i. e., the shortest baselines in order to compare the polarization pattern with other ALMA bands with the same angular resolution). We use only visibilities from the longest baselines (baselines longer than 4.6 km), which correspond to a *uv* distance larger than 1500 kλ (where λ = 3.1 mm, the wavelength of these observations), to resolve the compact scales. The resulting resolution is 84 mas (~ 14 au) with Briggs weighting robust factor of 1. The final continuum map has a noise level of 10 µJy beam$^{-1}$.
 The VLA observations were carried out in the A configuration for several runs between October and December 2016 (VLA project 16B-290). We observed four adjacent frequency bands, X (10.00 GHz), Ku (15.08 GHz), K (22.20 GHz) and Ka (34.50 GHz). The calibration was done using the VLA CASA pipeline (*21*). Images were produced with a robust weighting of -1.0, 0.5 and 1.0 to the Ku, K and Ka visibilities, respectively. These values yielded a synthesized beam of 173 x 73 mas (with a position angle of PA=19º) for the Ku band map, 173 x 73 mas (PA=10°) for the K band and 140 x 60 mas (PA=-5°) for the Ka band. The noise achieved is 19, 9 and 8 μJy beam$^{-1}$ for the Ku, K and Ka maps, respectively. The K and Ka maps resolve well the two radio sources, [BHB2007] 11A and [BHB2007] 11B. A Gaussian model was fitted to each source independently. The Ku map only resolves partially the two sources. To obtain the flux of the two sources, we fitted a Gaussian model by fixing the position of [BHB2007] 11A and [BHB2007] 11B (using the values from the Ka map). The resulting fluxes of the fits are shown in Table S1. To estimate the total flux within the circumbinary disk, maps in all bands were produced using Natural weighting to the visibilities, which yields an angular resolution similar to previous ALMA maps (*8, 11*). For Figure 3, we chose a center frequency of 32.5 GHz and produced an image with a robust weighting of 2.0, which yields a beam size of 168 x 92 mas (PA=-5°) and resolves the radio jet associated with source [BHB2007] 11A.

Spectral Energy Distribution of the circumbinary disk and protobinary components

From the flux densities reported in Table S1, we computed the spectral index $\alpha$ of the sources assuming that the flux varies as

$$S_\nu \propto \nu^\alpha, \tag{S1}$$

where $\nu$ is the band frequency. For the two sources, two different spectral indices were used to account for the different behavior at centimetric and millimetric wavelengths. For the circumbinary disk, the spectral index is consistent with optically thin free-free emission at the centimetric (VLA) range ($\alpha = -0.1$) and dust thermal emission at the millimetric (ALMA) bands ($\alpha = 3.0$). The ALMA spectral index is also reported in (*11*). The spectral indexes derived in the high resolution maps, where the protobinary system is resolved, are somewhat different. The spectral index at low frequencies is consistent with partially optically thick free-free emission ($\alpha = 0.6 \sim 0.7$) while the ALMA bands, as the previous case, have a steeper spectral index ($\alpha = 2.2 \sim 2.8$) that indicates optically thick dust thermal emission. Fitting the flux density $S_\nu$ as a function of the VLA and ALMA frequencies $\nu$ gives the following dependencies

$$[\text{BHB2007] 11A:} \quad S_\nu \text{ (mJy)} = 0.137 \, [\nu/10 \text{ GHz}]^{0.65} + 5.57 \, [\nu/250 \text{ GHz}]^{2.82} \tag{S2}$$

$$[\text{BHB2007] 11B:} \quad S_\nu \text{ (mJy)} = 0.0648 \, [\nu/10 \text{ GHz}]^{0.66} + 6.63 \, [\nu/250 \text{ GHz}]^{2.17} \tag{S3}$$

$$[\text{BHB2007]11 disk:} \quad S_\nu \text{ (mJy)} = 0.68 \, [\nu/10 \text{ GHz}]^{-0.1} + 178 \, [\nu/226 \text{ GHz}]^{3.0}. \tag{S4}$$

The fitted curves corresponding to equations (S2)-(S4) are shown in Figure S1.

Disk mass of the individual protostars of the binary system

The ALMA continuum data has contribution mostly from the dust thermal emission. We measured the flux densities, $S_\nu$, for each source component in the binary system. The ALMA data at 225.3 GHz has the highest angular resolution, producing a better contrast between the emission from the circumstellar disks and the emission from the circumbinary filaments.

Therefore, we use this map to fit a Gaussian model to determine the deconvolved beam sizes of the circumstellar disks. The inclinations of the normal to the disk planes with respect to the line-of-sight, determined from the ratio between their minor to major axis of the sources, are 43° ± 10° for [BHB2007] 11A and 41° ± 6° for [BHB2007] 11B. The integrated fluxes were determined from the Gaussian model over an intensity contour 50 times the noise of the map (19 µJy beam$^{-1}$). Table S1 summarizes the source properties.

The observed intensity of the dust emission at a particular frequency, $I_\nu$, can be described as a modified blackbody law, $B_\nu$, that depends on the dust temperature, $T_{dust}$, and a frequency-dependent optical depth, $\tau_\nu$, as

$$I_\nu = B_\nu(T_{dust})(1 - e^{-\tau_\nu}) = \frac{2h\nu^3}{c^2} \frac{1}{\exp(h\nu/kT_{dust})-1}(1 - e^{-\tau_\nu}), \tag{S5}$$

where $\nu$ is the observed frequency, $h$ is the Planck constant, $k$ is the Boltzmann constant, $c$ is the light speed and $\tau_\nu$ is defined as the integration of the dust opacity, $\kappa_\nu$, times the source volume density, $\rho$, along the line-of-sight, $\int \kappa_\nu \rho\, dl$, where $dl$ is the infinitesimal length scale of a source. In the Rayleigh-Jeans approximation, the brightness temperature, $T_b$, of a source is related to the observed peak intensity as

$$I_\nu = \frac{2k\nu^2}{c^2} T_b, \tag{S6}$$

where $T_b$ is related to the dust temperature, $T_{dust}$, through optical depth attenuation term as $T_{dust} = T_b/(1 - e^{-\tau_\nu})$. From the peak fluxes measured in our 225.3 GHz data, we find that $T_b$ = 35 ± 5 K and 62 ± 9 K for [BHB2007] 11A and [BHB2007] 11B, respectively (we removed a small but non-negligible contamination of free-free emission from the sources emission at 225.3 GHz, see below).

The flux density, $S_\nu$, of a source is defined as the integral of equation (S5) over the source angular size, $\Omega_{source}$, and can be written as

$$S_\nu = B_\nu(T_{dust})(1 - e^{-\tau_\nu})\Omega_{source} = B_\nu(T_{dust})(1 - e^{-\tau_\nu})\frac{A}{D^2}, \tag{S7}$$

where $A$ is the source emitting area (determined from the source projected size in Table S1) and $D$ is the distance to the source. Knowing that $M = A \times \int \rho\, dl$ and the definition of $\tau_\nu$, the equation above can be rewritten as

$$M_{d+g} = S_\nu \tau_\nu D^2 [B_\nu(T_{dust})(1 - e^{-\tau_\nu})\kappa_\nu]^{-1}, \tag{S8}$$

where $M_{d+g}$ is the circumstellar dust and gas mass. The flux density is obtained by the Spectral Energy Distribution (SED) analysis (equations S2 and S3). At 250 GHz, $S_{[BHB2007]\,11A}$ = 5.57 mJy and $S_{[BHB2007]\,11B}$ = 6.63 mJy. The largest source of uncertainty in Equation S8 is the dust opacity, $\kappa_\nu$, due to our lack of knowledge of the grain properties such as composition and dust size distribution in each disk. While dust models predict opacities of 2 to 3 cm$^2$ g$^{-1}$ from grains with maximum size smaller than 100 µm in protoplanetary disks (*22*), in younger objects such as [BHB2007] 11, which has an estimated age of 0.1 to 0.2 Myr (*7*), $\kappa_{250GHz}$ is heavily dependent on the dust composition and varies roughly between 0.8 and 1.0 cm$^2$ g$^{-1}$ (*23*). However, because we are observing dense gas, we should expect some increase in the opacity due to grain growth. We therefore assume a dust opacity $\kappa_{250\mathrm{GHz}}$ per gram of dust of 1.7 cm$^2$ g$^{-1}$ as a compromise that accounts for young disk chemistry and protoplanetary disk grain sizes. This value is also consistent with (24), who proposes that $\kappa_\nu = 0.02(1\ \mathrm{mm}/\lambda)^\beta = 1.67$ cm$^2$ g$^{-1}$ for a dust opacity index $\beta = 1$, which is the case when dust grains are expected to be larger than in the interstellar medium (24), and $\lambda = 1.2$ mm (corresponding wavelength for $\nu = 250$ GHz). In addition, for a young disk, we adopt the canonical gas-to-dust ratio of 100 (25), although more evolved disks are expected to have a lower ratio (26). The assumed dust opacity per total mass is thus 0.017 cm$^2$ g$^{-1}$. Theoretical modeling of dust evolution in disks suggests that dust temperatures at a few au disk radii are proportional to the stellar luminosity (27). From the [BHB2007] 11 bolometric luminosity reported in (7), we assume that each component of the protobinary has a luminosity of 1.75 L$_\odot$[2]. This value is consistent with a dust temperature of ~ 110 K at 3 au disk radius (27), which is a factor 2 to 3 higher than the brightness temperatures of each source in an optically thin regime. From equation (S8), this sets a lower limit on their disks masses of ~ 0.3 M$_{\mathrm{jup}}$. If the continuum emission is optically thick ($\tau > 1$), as suggested by the spectral indexes in equations (S2) and (S3), the disk temperatures are closer to the brightness temperatures and the disk masses should be of the order of the Jupiter mass or larger. For instance, for $\tau = 5$, the disk masses of [BHB2007] 11A and [BHB2007] 11B are ~ 4 M$_{\mathrm{Jup}}$ and ~ 3 M$_{\mathrm{Jup}}$, respectively.

Filament masses

For the filaments, the mean brightness temperature along their structure is 14 ± 2 K, peaking at 31 ± 5 K near [BHB2007] 11B. After subtracting the contribution from the protobinaries, we measured the filaments flux density for pixels brighter than 15 times the noise of the dust map (0.019 mJy beam$^{-1}$) as 0.14 ± 0.02 Jy. The peak intensity is 1.8 mJy beam$^{-1}$ and the mean intensity over the 15 × noise level is 0.79 mJy beam$^{-1}$. The aproximate total length determined from the brightest pixels along the spine of the filament within the 15 × noise level is 392 au, and the mean width is 12 au (Figure S2A). The filaments are expected to be cooler than the disk, so we assume a dust temperature range of 40 to 100 K. In the case of optically thin emission, this suggests a filament mass (equation S8) between ~ 9 M$_{\mathrm{Jup}}$ and 29 M$_{\mathrm{Jup}}$. This is equivalent to a mass per unit length of ~ 5-14 M$_\odot$ pc$^{-1}$, which is smaller than the critical mass per unit length for an isothermal cylinder, $M_{\mathrm{line,crit}} \approx 16 \left(T_{\mathrm{gas}}/10\ \mathrm{K}\right)$ M$_\odot$ pc$^{-1}$ ~ 16 M$_\odot$ pc$^{-1}$ (assuming $T_{\mathrm{gas}} = 10$ K).

Kinematics and dynamical mass determination from molecular line emission

Using the molecular line data reported in (8, 11), we performed two-dimensional Gaussian fits on the molecular emission on each velocity channel to determine the peak intensity. Then we computed the uncertainty on the source position as $\frac{\Theta}{\mathrm{SNR}}$, where $\Theta$ is the angular resolution (the synthesized beam) of the molecular map and SNR is the signal-to-noise ratio, i. e, the ratio of the peak intensity to the map noise, thus the uncertainty in the position is inversely proportional to the SNR in each velocity channel. We computed peak positions for the formaldehyde (H$_2$CO) line at 218.76 GHz (H$_2$CO $3_{2,1} - 2_{2,0}$, whose upper energy level, $E_U$, is 68.1 K) and the CO (2 – 1) line, whose results are discussed in the main text. All components have positional accuracies better than 0.05″ (we used a threshold SNR of 5).

---

[2] The luminosity of each protostar is assumed to be half of the value reported in (7), and it is also scaled for the Pipe Nebula distance of 163 pc (10) (the assumed distance for the Pipe Nebula in (7) is 130 pc).

The H$_2$CO data indicate that gas at the systemic, local standard of rest velocity (lsr) of [BHB2007] 11, $v_{lsr} \sim 3.6$ km s$^{-1}$, and the velocity gradient toward their redshifted ($v_{lsr} > 7.3$ km s$^{-1}$) and blueshifted ($v_{lsr} < 0.3$ km s$^{-1}$) components are primarily associated with the rotation of the circumbinary disk (Figure S3A). The intensity-weighted mean velocity of this transition is consistent with disk rotation (*11*). Using the molecular line data cube, we generated a position-velocity diagram and, assuming the velocity varies with the distance, $d$, to the disk center as $v \propto \sqrt{M_{dyn}/d}$ where M$_{dyn}$ is the dynamical mass of the system, we fitted a Keplerian model to emission brighter than 5 times the noise level of the PV diagram (~3 mJy beam$^{-1}$), following the method proposed by (*28*). The Keplerian model fit resulted in a dynamical mass (deprojected from the line-of-sight) of 2.25 ± 0.13 M$_\odot$, which represents an upper limit to the combined mass of the protostars.

We can estimate the mass accretion rate from the circumbinary disk into the circumstellar disk of [BHB2007] 11B by assuming that the material at the far end side of the filament will fall freely into the star. The free-fall time is given by

$$t_{ff} = \frac{\pi}{2}\left(\frac{l_{inf}^3}{2GM_*}\right)^{\frac{1}{2}} \tag{S9}$$

where $l_{inf}$ is the path followed by the falling material and $M_*$ is the central object mass. We assume that the central star in [BHB2007] 11B is slightly less massive than half of the combined mass computed above, that is $M_* \approx 1.0$ M$_\odot$, and that the filament's length, $l_{inf}$, along the CO blueshifted peaks (Figure 2A) is ~ 28 au with a mass of $m \approx 1.4 \times 10^{-4}$ M$_\odot$. Considering two streams accreting into the star, these values provide a mass accretion rate of ~ $1.1 \times 10^{-5}$ M$_\odot$/year, which is consistent with accretion rates reported in previous works (*16*) and with the infall velocities of CO line. This is a rough estimate of the value of the mass accretion rate because, in addition to the uncertainties of the quantities used in this calculation, we assumed that the falling material starts its movement from rest and follows a straight line.

Dynamical state of the circumbinary disk

We have estimated the global Toomre parameter $Q$ for the circumbinary disk to assess the disk hydrodynamical stability. The Toomre parameter indicates if the disk is massive enough to be gravitationally unstable ($Q \approx 1$ or less) or if it is stable against gravity (through resisting forces such as thermal pressure) ($Q \gg 1$). In rotationally supported disks, the Toomre factor can be approximated as (*14, 15*)

$$Q \approx 2\frac{M_\star}{M_{disk}}\frac{H}{R} \tag{S10}$$

where $M_\star$ is the combined protostellar masses, $M_{\text{disk}}$ is the circumbinary disk mass, $H$ is the disk scale height that depends on the sound speed, $c_s$, of the disk and the Keplerian angular velocity, $\Omega$, at the radius $R$. Using equation S8 with the parameters reported in (8), the circumbinary disk mass is $0.08 \pm 0.03$ M$_\odot$. The combined protostellar mass computed above, $2.25 \pm 0.13$ M$_\odot$, was determined from H$_2$CO emission confined in a radius $R$ of 90 au within the circumbinary disk. The angular velocity $\Omega$ at this radius, $\sim 3.5 \times 10^{-10}$ rad s$^{-1}$, is also estimated from the H$_2$CO emission. Taking $c_s$ as $\sim 0.33$ km s$^{-1}$ (for a mean disk temperature of 30 K), the Toomre parameter $Q$ is $\sim 3.9$. It is thus unlikely that the filaments are the result of fragmentation of the disk, especially with no signs of formation of companion objects along their structures. This reinforces our conclusions that the filaments are accretion streamers and also possibly the result of tidal interactions between the binary and the material in the circumbinary disk.

Ionized mass loss rate from each protobinary component

The nearly flat spectral index observed at centimetric wavelengths for each component is consistent with a thermal radio jet (19). From the VLA fluxes in Table S1, we can derive the ionized mass loss rate of each object. Using the best fitting spectral index from the centimetric bands (blue and green curves in Figure S1), we can determine the contribution of free-free emission in each of the VLA frequencies (15, 22 and 34 GHz). As shown in (11), the 34 GHz emission is well resolved between [BHB2007] 11A and [BHB2007] 11B. Therefore, in order to avoid mutual contamination, we compute the free-free contribution at this band. From equations (S2) and (S3), we have

$$S_{\text{[BHB2007] 11A}} = 0.137 \left(34.5/10 \text{ GHz}\right)^{0.65} = 0.306 \text{ mJy} \tag{S11}$$

$$S_{\text{[BHB2007] 11B}} = 0.0648 \left(34.5/10 \text{ GHz}\right)^{0.66} = 0.147 \text{ mJy}, \tag{S12}$$

where $S_{\text{[BHB2007] 11A}}$ and $S_{\text{[BHB2007] 11B}}$ are the free-free emission fluxes in each protostar. We use the model presented by (19) for free-free emission from an ionized jet to compute the ionized mass loss rate. Assuming a radio jet with constant opening angle, the ionized mass loss rate $\dot{M}_{\text{ion}}$ is given in observational units (29) as

$$\dot{M}_{\text{ion}} = 0.0796 \left[\left(\frac{S_{34 \text{ GHz}}}{\text{mJy}}\right)\left(\frac{\theta}{\text{rad}}\right)\right]^{0.75} \left(\frac{v}{200 \text{ km s}^{-1}}\right) (\sin i)^{-0.25} \left(\frac{d}{\text{kpc}}\right)^{1.5} \left(\frac{T}{10^4 \text{ K}}\right)^{-0.075}, \tag{S13}$$

where $S_{34 \text{ GHz}}$ is the free-free contribution computed above, $\theta$ is the injection opening angle of the jet, $v$ is the jet velocity, $i$ is the inclination of the sources, $d$ is their distance and $T$ is the electron temperature. The result of the above equation is given in units of $10^{-6}$ M$_\odot$ year$^{-1}$. We adopted $\theta = 10°$ since the usually high collimation of jets allow for a small opening angle, $v = 200$ km s$^{-1}$ as typical jet velocity for low-mass objects, the inclination, $i$, determined from the source sizes (see main text) and T = $10^4$ K is usually adopted (29). Despite the flux calibration uncertainties at centimetric-wavelength observations being normally lower than millimetric ones, we assume a conservative uncertainty for the VLA fluxes to be similar to the ALMA data, 15%. The ionized mass loss rate is thus $(6.4 \pm 0.7) \times 10^{-10}$ $M_\odot$ year$^{-1}$ for [BHB2007] 11A and $(3.7 \pm 0.4) \times 10^{-10}$ $M_\odot$ year$^{-1}$ for [BHB2007] 11B.

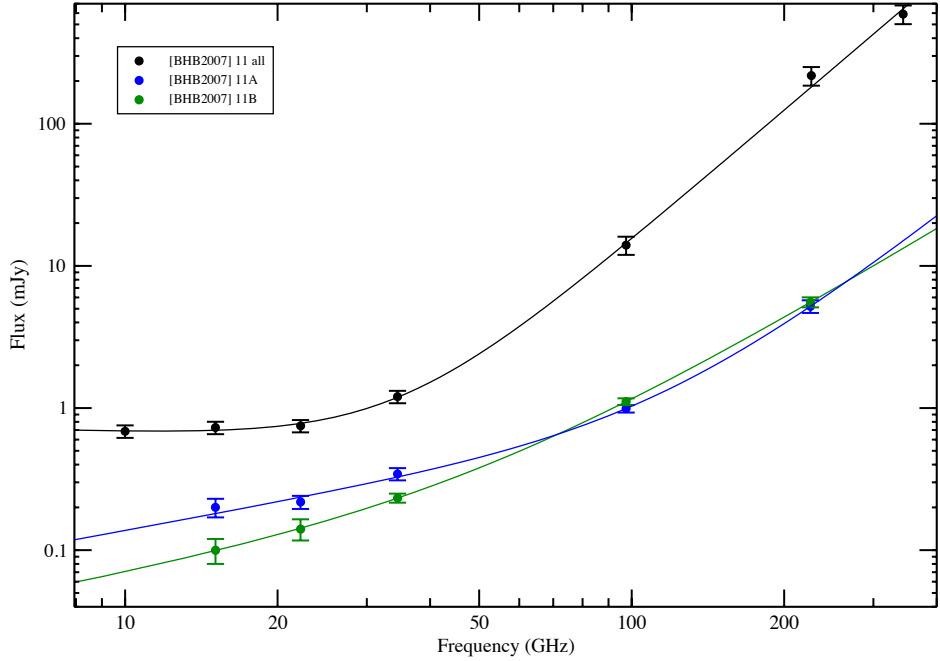

**Fig. S1. Flux density as a function of the frequency and spectral index derivation**. The emission from the circumbinary disk and the emission from each binary component are shown whenever they are resolved. The lines represent the best fitting model of the flux density as a function of frequency for the circumbinary disk (black line), [BHB2007] 11A (blue line) and [BHB2007] 11B (green line). The fitting functions are mathematically described in equations (S2)-(S3). For the circumbinary disk, the ALMA fluxes are retrieved from previous lower resolution continuum maps (*11*). For the VLA, we performed imaging with a Gaussian taper in the visibility plane in order to obtain a similar angular resolution as the previous ALMA maps (*8, 11*). The VLA fluxes from the circumbinary disk are $0.68 \pm 0.07$ mJy (10 GHz), $0.73 \pm 0.07$ mJy (15 GHz), $0.75 \pm 0.08$ mJy (22 GHz) and $1.20 \pm 0.12$ (34 GHz). At 10 GHz, the individual emission from each binary component is not resolved. The Band 3 ALMA fluxes were determined using a filter in the *uv* visibilities (*uv* distance larger than 1500 k$\lambda$) to avoid contamination from the dust streamers. The fluxes of the individual objects are listed in Table S1.

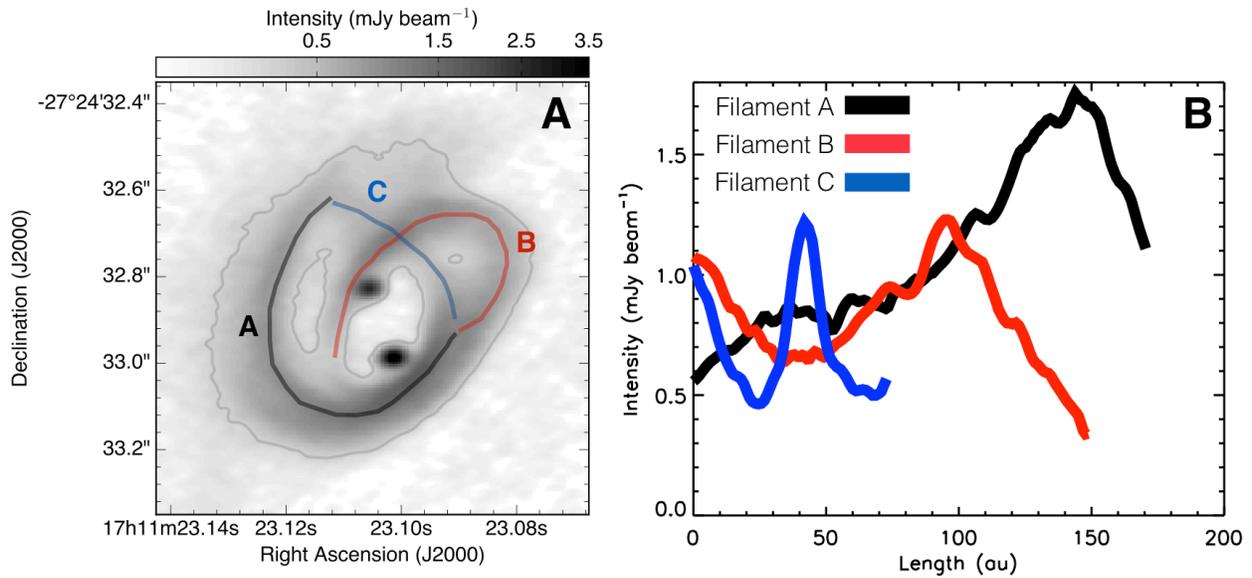

**Fig. S2**: **Peak intensity of brightest filaments**. 1.3 mm continuum emission map showing an intensity contour level 15 times the noise level of the map (~ 19 µJy beam$^{-1}$). The colored lines show the bright pixels along the spine of the filaments to highlight the main structures (the distinct colors do not necessarily mean that the filaments are not part of a single structure). Panel B shows the intensity cut along the spine of each filament. The intensity peak in filament C is due to the region where filaments B and C intercept (in projection).

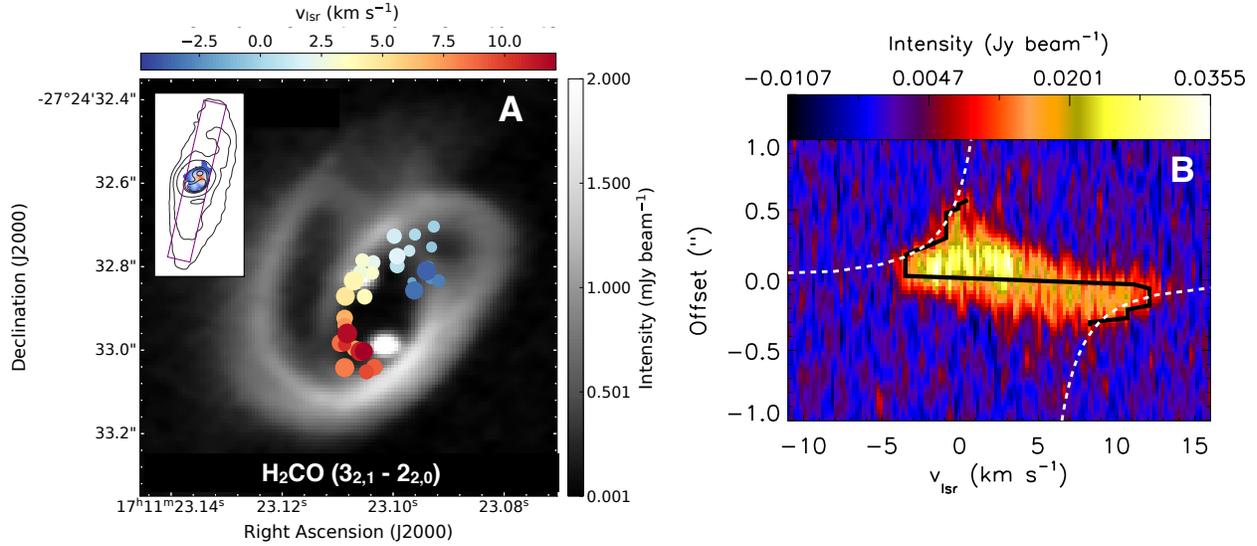

**Fig. S3. Kinematics revealed by H$_2$CO data**. (A) positions of the velocity components of the H$_2$CO emission with red (blue) circles indicating the components that are redshifted (blueshifted) with respect to the local standard of rest velocity of the source ($v_{sr}$ ~ 3.6 km s$^{-1}$, shown as yellow circles). The H$_2$CO emission peaks are drawn over the continuum image. The separation between velocity components is 0.4 km s$^{-1}$ for H$_2$CO (and 0.35 km s$^{-1}$ for CO, see Figure 2A). The inset shows the lower resolution continuum contours of [BHB2007] 11 (also in Figure 1B) with the integrated intensity map of the H$_2$CO emission for pixels brighter than 5 times the noise level (the map peaks at 0.3 Jy beam$^{-1}$ km s$^{-1}$). The rectangle indicates the rectangular region used to determine the position-velocity (PV) diagram of the H$_2$CO emission. (B) PV diagram of the H$_2$CO emission from a cut with a position angle of 167° (counted from North to East), which is perpendicular to the outflow direction reported in (*8*), and widh of ~ 0.57″, which is equivalent to the size H$_2$CO emission. The black line shows emission 5 times brighter than the noise and the white dashed line shows our Keplerian model fitted to the data leading to a combined mass of 2.25 ± 0.13 M$_\odot$.

**Table S1.**

Flux density in mJy for each component of the protobinary system in each observed frequency and source sizes (deconvolved from the synthesized beam) obtained from a Gaussian fit on the 225.3 GHz map. The fluxes at 15, 22 and 34 GHz were obtained from the VLA data, while the fluxes at 97.5 and 225.3 GHz were obtained from the ALMA data.

| Sources [BHB2007] | Flux (15 GHz) | Flux (22 GHz) | Flux (34 GHz) | Flux (97.5 GHz) | Flux (225.3 GHz) | Deconvolved size (au) | Position angle (°) |
|---|---|---|---|---|---|---|---|
| 11a | 0.20 ± 0.03 | 0.22 ± 0.02 | 0.34 ± 0.03 | 0.99 ± 0.06 | 5.20 ± 0.53 | 7.3 × 5.4 (±1.3) (±1.1) | 79 ± 89 |
| 11b | 0.10 ± 0.02 | 0.14 ± 0.02 | 0.23 ± 0.02 | 1.12 ± 0.06 | 5.62 ± 0.45 | 4.9 × 3.7 (±1.1) (±1.0) | 78 ± 83 |